\documentclass[aps,prb,twocolumn,groupedaddress,epsfig,showpacs]{revtex4}
\usepackage{graphicx}
\usepackage{amsmath}

\begin{document}

\title{Far-infrared absorption of self-assembled semiconductor rings\ddag}

\author{J. Planelles}
\email{planelle@exp.uji.es}
\author{J.I. Climente}
\affiliation{Departament de Ci\`encies Experimentals, UJI, Box 224, E-12080 Castell\'o, Spain}
\date{\today}

\begin{abstract}

We report a theoretical description of far-infrared spectroscopy experiments on self-assembled quantum rings in a magnetic field [A. Lorke \emph{et al.}, Phys. Rev. Lett. {\bf 84}, 2223 (2000)] which, for the first time, accounts for the full set of experimental resonances. In our calculations we use a 3D effective-mass model with a realistic finite step-like confinement potential, including strain and Coulomb effects. We assume a bimodal distribution of ring sizes.

\vspace{1cm}
\noindent {\bf \ddag We dedicate this paper to Josef Paldus in celebration of his 70th anniversary and his many outstanding  contributions to quantum chemistry.}

\vspace{1cm}
\end{abstract}


\maketitle

\section{Introduction}

Quantum rings have received a great deal of attention from researchers in the condensed matter field, over the last two decades.\cite{ButtikerPL,AronovRMP,ChakraPRB,WendlerPSSb,FuhrerNAT}
It is mainly the magnetic properties of these systems that make them so interesting.
When a quantum ring is pierced by a magnetic field perpendicular to the plane of the ring, the Aharonov-Bohm effect\cite{AharonovPR} leads to a persistent current of the charge as well as to oscillations of its energy.\cite{OlariuRMP}
Experimental evidence for Aharonov-Bohm oscillations is available from metallic and semiconductor rings in the mesoscopic regime.\cite{ChandrasekharPRL,FuhrerNAT}
However, mesoscopic rings may be subject to electron-electron, electron-impurity and electron-phonon scattering.
Thus, the recent realization of nanoscopic disorder-free few-electron quantum rings is acknowledged as a major developement in the low-dimensional domain where new physics driven by confinement, electron correlations and the influence of an external applied field can be explored.\cite{Chakra_book}

Nanoscopic rings can be synthesized either by means of self-assembly\cite{GarciaAPLRazAPL} or litographic techniques.\cite{BayerPRL}
One advantage of self-assembled structures over their litographic counterparts is a superior optical quality, which makes them especially attractive for device applications.\cite{SkolnickPE}
Although atomic force micrographs of self-assembled rings evidence their ringlike geometry,\cite{GarciaAPLRazAPL} such images are taken before the rings sample is covered with matrix material. Since embedding these structures in a semiconductor matrix is essential for practical applications, spectroscopy experiments were performed on covered samples to confirm whether the ringlike geometry is still preserved or not.\cite{WarburtonNAT,PettersonPE,HaftPE,LorkePRL,EmperadorPRB} 
Far-infrared (FIR) absorption spectroscopy on a macroscopic number of self-assembled rings, each of which charged on average with one\cite{EmperadorPRB} and two\cite{LorkePRL} electrons, was measured as a function of an external magnetic field.
Characteristic spectral features at about $B=8$ T were attributed to the change, from $0$ to $-1$, of the ground state  z-projection of angular momentum.\cite{LorkePRL,EmperadorPRB,HuPRB}
This interpretation of the experiments relies on two-dimensional effective mass models with a parabolic-like confinement potential (the potential typically used to investigate mesoscopic rings\cite{ChakraPRB}).
Indeed, these models yield reasonable agreement with most of the experimental data, but they also involve a few issues not completely understood, namely:

(i) Different \emph{characteristic frequencies} of the confinement potential are needed to fit the corresponding experiment if the rings contain one or two electrons, even though the ring sample is actually the same.\cite{EmperadorPRB,HuPRB}

(ii) An \emph{effective radius} $R=14$ nm is needed to fit the experiment.\cite{LorkePRL,EmperadorPRB,HuPRB} This is somewhat surprising because atomic force micrographs show the inner radius to be $R_{in}=10-15$ nm and the outer radius $R_{out} \approx 60$ nm.\cite{LorkePRL,BarkerPRB}

(iii) The calculated relative intensities of the low-lying and high-lying sets of resonances differ by at least one order of magnitude,\cite{EmperadorPRB,HuPRB} whereas, experimentally, they are found to have similar oscillator strength.\cite{LorkePRL}

(iv) The highest energy resonances which are calculated in the presence of a magnetic field overestimate the energy position of the corresponding experimental peaks (marked with dots in Figures 3 and 6 of Ref.\onlinecite{EmperadorPRB}).\cite{EmperadorPRB,HuPRB}

(v) A few experimental resonances (marked with crosses in Figures 3 and 6 of Ref.\onlinecite{EmperadorPRB}) cannot be explained.\cite{EmperadorPRB,HuPRB}

The first and second issues are inherent shortcomings of two-dimensional models with parabolic-like confinement potential, which require precise knowledge on the energy spectrum beforehand to fit several parameters.
The third and fourth issues were overcome in a later work by Puente and Serra using a two-dimensional model with an improved form of the parabolic-like potential barrier for the inner radius of the ring.\cite{PuentePRB}
However, their model brought about a new interpretation of the FIR experiments suggesting that a mixture of high- and low-barrier rings must be contained in the sample of Ref.\onlinecite{LorkePRL}.
Moreover, the change in transition energy at $B=8$ T was no longer attributed to an Aharonov-Bohm oscillation but to a crossing between the energy levels of the two different types of quantum ring.
This hypothesis of a bimodal distribution of ring sizes agrees with recent observations on near-infrared spectroscopy of self-assembled rings.\cite{HaftPE,WarburtonICPS,ClimentePRB}
Ref.\onlinecite{PuentePRB} calculations also received strong support from our recent work.\cite{ClimentePRBee} In this work, by describing a two-electron self-assembled ring with a truly, fittings-free, three-dimensional model, we obtained results 
which are very similar to their predictions for high-barrier rings.

In this paper, we calculate the energy levels and FIR absorption spectra of one- and two-electron  InAs/GaAs dots and rings with two different inner radii. The model used is the same as in Ref.\onlinecite{ClimentePRBee}, which includes strain and Coulomb effects, as well as, a realistic finite confinement potential to describe the semiconductor heterostructure interface. Our results show that the combined absorption spectra of the two rings agree qualitatively well with the experimental data and overcome all the aforementioned shortcomings of two-dimensional models. To our knowledge, this is the first theoretical description which is able to account for all the experimental FIR resonances of Ref.\onlinecite{LorkePRL}, including those marked with crosses. 

\section{Theory and calculation methods}
The one-band effective mass Hamiltonian for the electron states, including a magnetic field perpendicular to the ring plane, can be written in atomic units as

\begin{multline}
{\cal H}_e= \left( -\frac{1}{2} \nabla \left(\frac{1}{m^*(E_{n,m};\rho,z)}\,\nabla \right) + \frac{(B\,\rho)^2}{8m^*(E_{n,m};\rho,z)} \right.\\
+\frac{B\,m}{2m^*(E_{n,m};\rho,z)} +\frac{1}{2}\,\mu_B\,g(E_{n,m};\rho,z)\,B\,\sigma\\
\biggl. +V_c(\rho,z)+a_c\,\varepsilon_{hyd}(\rho,z) \biggr),
\label{eq1}
\end{multline}

\noindent where $m=0,\pm 1,\pm 2,\ldots$ is the quantum number of the projection of the angular momentum onto the magnetic field ($B$) axis, $n$ is the main quantum number, $V_c (\rho,z)$ is the finite confinement potential corresponding to the geometries shown in Figure 1, and $m^*(E_{n,m};\rho,z)$ and $g(E_{n,m};\rho,z)$ stand for
 the energy- and position-dependent mass and Land\'e factor, respectively.\cite{VoskoboynikovPRB} $a_c$ denotes the
hydrostatic deformation potential for the conduction band, and $\varepsilon_{hyd}$ is the hydrostatic strain, which
we calculate within the framework of the isotropic elastic theory.\cite{DownesJAP,DaviesJAP}
It should be underlined that $V_c$ must be a step-like, finite confinement potential
in order to achieve a realistic description of the effect of the inner radius and the magnetic field
penetration into the ring region.\cite{VoskoboynikovPE,LiJAP}
A configuration interaction procedure is used to calculate the two-electron eigenstates and eigenenergies.
The two-electron states can be labeled by the $z$ projection of the total angular momentum $M=m_1+m_2$, total spin
$S=\sigma_1+\sigma_2$, and main quantum number N. 
The optical absorption intensities for intraband transitions between electron states are calculated within the electronic dipole approximation.\cite{Loehr_book} We assume non-polarized light, although most of the intensity arises from the in-plane light components. We also assume $T=0$ K, and therefore only transitions from the ground state are calculated. In order to obtain smooth spectra, the transition probabilities are represented employing Lorentzian curves of half-width $\Gamma=0.5$ meV. Further details about the theoretical model are given in Ref.\onlinecite{ClimentePRBee}.

\section{Results and discussion}
We investigate three self-assembled InAs quantum structures embedded in a GaAs matrix.
Their cross-sections on the $(\rho,z)$ plane are represented in Figure 1.

\begin{figure}[h]
\includegraphics{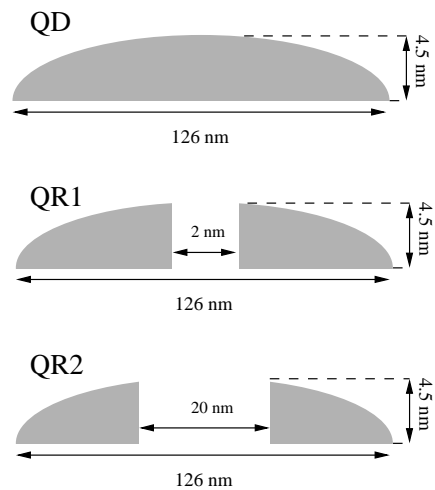}
\caption{Schematic cross-section of the three structures under study.}
\label{Figure1}
\end{figure}

The first structure (QD) is a compact quantum dot, the second one (QR1) is a quantum ring with a small inner radius of $R_{in}=1$ nm and the third structure (QR2) is a quantum ring with an inner radius of the size measured by atomic force microscopy, $R_{in}=10$ nm.
All three structures have an (outer) radius of 63 nm and a height of $4.5$ nm, which are close to the experimental dimensions observed in uncovered self-assembled rings.\cite{LorkePRL,BarkerPRB}
The three structures can be seen as different stages of developement of a quantum ring.
The dot is lens-shaped and the shape of the rings is a cut torus with sheer inner wall.
Experimentally, it has not been established if the quantum rings are made of pure InAs or an InGaAs alloy.
However, we assume pure InAs composition for all three structures and use the same material parameters as in Ref.\onlinecite{ClimentePRBee}.
An InGaAs alloy is expected to yield similar results
to those we predict here because the increased electron effective mass due to the presence of Ga would be compensated by
the weaker strain effects, which in turn would lead to a smaller strain-induced increase of the effective mass.

Equation (\ref{eq1}) is integrated numerically by employing finite differences in a two-dimensional grid $(\rho,z)$.
The configuration interaction calculations include all the single-particle states up to 35 meV away from the ground
state. We have determined that the use of larger basis sets does not significantly change the low-lying two-electron states within the range of the magnetic field that is studied.

\subsection{Single-electron systems}

Figure 2 illustrates the monoelectronic energy levels of QD, QR1 and QR2 vs.\ a magnetic field. Solid and dotted lines are used for spin up and spin down levels, respectively.
As we already discussed for very similar structures,\cite{ClimenteXXmag} 
the presence of the hole in the ring
significantly reduces the energy spacing
between consecutive azimuthal levels ($m=0,\pm 1,\pm 2,\ldots$) at $B=0$.
As a result, changes in the $z$-component of
the ground state angular momentum of QR1 and QR2 take place in the magnetic field range
under study, 0-12 T. The ground state of QR1 undergoes a change in angular momentum from $m=0$ to $m=-1$ at about 11.6 T,
whereas QR2 changes from $m=0$ to $m=-1$ at about 1.9 T, from $m=-1$ to $m=-2$ at about 5.8 T, etc.
In contrast, no angular momentum changes occur in the ground state of the one-electron dot, where the low-lying levels
converge to the first Landau level without crossings.\cite{Chakra_book}
The oscillations in the ground state energy due to angular momentum changes are a manifestation of the Aharonov-Bohm
effect and may be used as evidence of ring geometry.

\begin{figure}[h]
\includegraphics{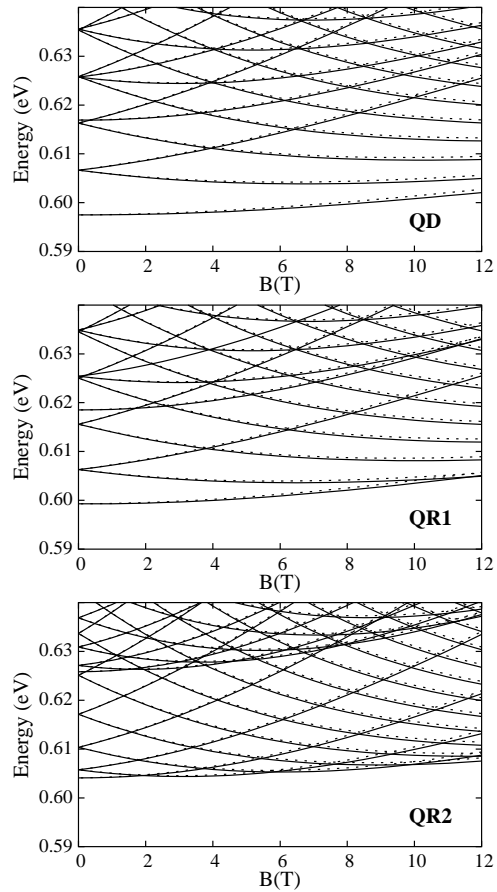}
\caption{Energy levels vs.\ magnetic field of one electron in QD (upper panel), QR1 (middle panel) and QR2 (lower panel).
Solid lines denote spin up and dotted lines spin down levels.}
\label{Figure2}
\end{figure}

Figure 3 shows the low-energy FIR absorption of one electron in QD, QR1 and QR2
for $B=0,0.5,1,1.5 \dots ,12$ T.
The intensities are displayed in arbitrary units and they are offset for clarity.

\begin{figure}[h]
\includegraphics{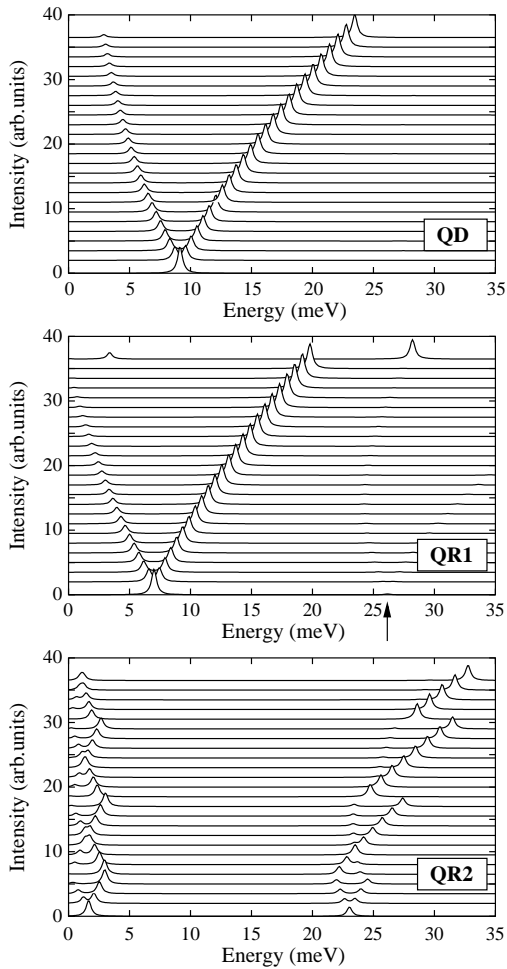}
\caption{FIR absorption of one electron in QD (upper panel), QR1 (middle panel) and QR2 (lower panel) at $T=0$ K.
The spectra are calculated for magnetic fields $B=0-12$ T in steps of $0.5$ T. The intensities are in arbitrary units and the curves have been offset for clarity. The arrow in QR1 points the position of the $\Delta n=1$ resonance.}
\label{Figure3}
\end{figure}

The spectra of QD show a single visible peak at $9.1$ meV for $B=0$. This peak stems from the $\Delta n=0,\,\Delta m=\pm 1$ transitions. When the magnetic field is switched on, the spectra split into two branches. One branch decreases in energy and intensity with the increasing external field. This branch is connected with the $\Delta m=-1$ transition. The other branch, connected with the $\Delta m=+1$ transition, increases in energy and remains intense even for strong magnetic fields.
The spectra of QR1 are similar to those of QD except for two important features. First, at $12.0$ T both the $\Delta m=1$ and the $\Delta m=-1$ branches are abruptly shifted toward higher energies. This is a consequence of the ground state change from $m=0$ to $m=-1$ which takes place at $11.6$ T. 
Second, a new peak stemming from the $\Delta n=1,\,\Delta m=\pm 1$ transitions arises at $26.1$ meV in the absence of a magnetic field. This peak, pointed by an arrow in Figure 3, is still very small (only $4\%$ of the $\Delta n=0$ peak intensity). However, it is an order of magnitude stronger than the corresponding transition resonance in the QD case.
Both features (the abrupt shift in resonance energies as the magnetic field increases and the presence of an intensity-enhanced $\Delta n=1$ resonance) are distinctive of ringlike geometry.
This is confirmed in view of the spectra of QR2, 
where the two peaks at $B=0$ are already of comparable intensity.
These two peaks split into $\Delta m=1$ and $\Delta m=-1$ branches in the presence of a field. The energies of the
 $\Delta n=0$ resonances follow a zig-zag course which reveals the underlying Aharonov-Bohm oscillations in the energy structure (see QR2 panel in Fig.\ 2). The energies of the most intense $\Delta n=1$ resonances also exhibit abrupt shifts in energy connected with Aharonov-Bohm oscillations in energy, but now they decrease the energy of the resonance. Unlike in the QD and QR1 cases, the rapid oscillations in the ground state of QR2 prevent the resonances from reaching energies between 5 and 20 meV. This energy gap seems to be characteristic of well-developed rings.\cite{PuentePRB}

A comparison with the experimental FIR absorption of one-electron self-assembled rings considering only QR2 would reproduce some experimental features while disregarding others. It has been suggested that the experimental sample may contain a mixture of well-developed rings plus partially-developed rings\cite{PuentePRB} or even large quantum dots which have not developed into rings.\cite{LorkePRL}
We study these possibilities by representing the combined spectra of QR1 and QR2 and comparing it with the one-electron experimental resonances.
The result is illustrated in Figure 4, where solid lines stand for the absorption of QR1 and dashed ones for that of QR2.
The experimental resonances are denoted by the same symbols as in Ref.\onlinecite{EmperadorPRB},
which show the way the resonances were originally grouped.
It should be mentioned here that in the experiments,\cite{EmperadorPRB,LorkePRL} an insufficient signal-to-noise ratio at low energies does not allow the detection of the resonances under $10$ meV.
It can be seen in Figure 4 that the dots are reasonably well described by QR2. Only the dot located at $B=6$ T significantly deviates from the corresponding calculated resonance. However, it is quite close to the calculated peak at $B=5.5$ T, so that one may easily achieve a better agreement if only the ground state change (from $m=-1$ to $m=-2$) we predict at $B=5.8$ T would be postponed to $B>6$ T.
The agreement of QR1 with the crosses and triangles is more questionable. The $\Delta m=1$ branch of QR1 lies in intermediate energies between those of the crosses and those of the triangles. Although the slope of this branch agrees with the slope of both types of experimental resonances, it is difficult to determine if the branch reproduces qualitatively the crosses or the triangles. In any event, it seems that only one of these two types of experimental resonances may be explained by the $\Delta m=1$ branch. 
A very similar picture to Figure 4 would arise by including QD in the mixture, since its absorption is very similar to that of QR1.
We will come back to this issue later in the paper.

\begin{figure}[h]
\includegraphics{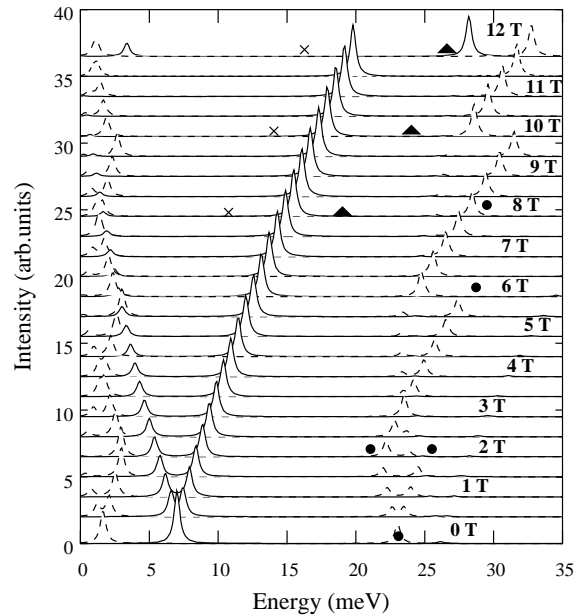}
\caption{Combined FIR absorption of one electron in QR1 and QR2 at $T=0$ K.
The spectra are calculated for magnetic fields $B=0-12$ T in steps of $0.5$ T. The intensities are in arbitrary units and the curves have been offset for clarity. 
Solid lines are used for QR1 and dashed lines for QR2. The symbols represent the experimental resonances.}
\label{Figure4}
\end{figure}

\subsection{Two-electron systems}

Next, we calculate the two-electron energy levels and FIR absorption of QD, QR1 and QR2.
The energy levels of the two-electron QD, QR1 and QR2 systems vs.\ a magnetic field are depicted in Figure 5. Only the levels which become ground state within the 0-12 T range are displayed. The quantum numbers $(M,S)$ of each level are also indicated (all the levels shown have $N=1$).

\begin{figure}[h]
\includegraphics{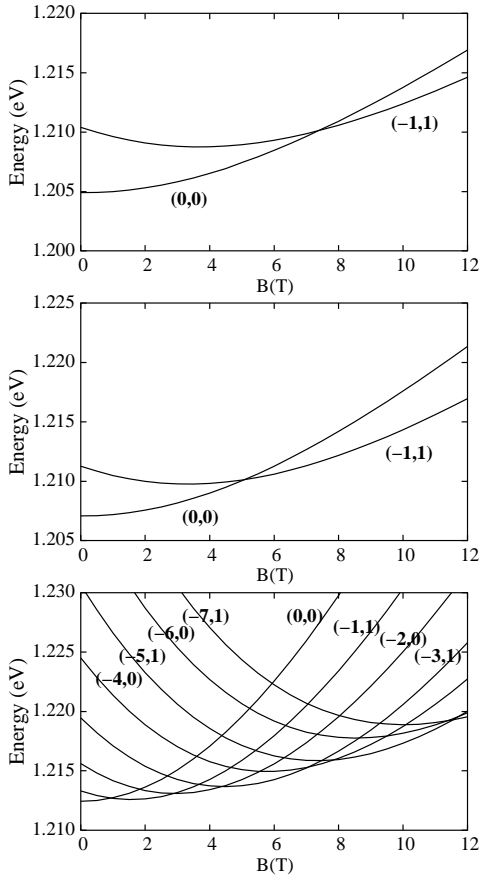}
\caption{Energy levels vs.\ magnetic field of two electrons in QD (upper panel), QR1 (middle panel) and QR2 (lower panel). Only levels that become the ground state in a given magnetic field window are shown. (M,S) labels denote the total angular momentum z-projection and total spin of each level.}
\label{Figure5}
\end{figure}

In the absence of a magnetic field, the ground state in QD is $(0,0)$ and remains so for magnetic fields as strong as $7.5$ T.\cite{spintransPRB} When $B>7.5$ T, the first spin-singlet-spin-triplet transition of the ground state takes place\cite{WagnerPRB} and it becomes $(-1,1)$.
In QR1, the spin-singlet-spin-triplet transition occurs at a weaker magnetic field than in the QD case. This is due to the hole in the ring, which has a strong effect on the magnetization even if it is small.\cite{ClimenteXXmag}
This first crossing in the ground state is found at a lower field than in the single-electron case (Figure 2). This is due to the direct and exchange Coulomb energies.\cite{ClimentePRBee}
The impact of the hole becomes dramatic for QR2, where up to six singlet-triplet crossings occur within the 0-12 T range.
One can also compare this with the single-electron QR2 case, where only three crossings occur in the ground state for the same magnetic field range. The increase in the number of crossings again is a reflection of the electron-electron interactions.

Figure 6 shows the low-energy FIR absorption of two electrons in QD, QR1 and QR2,
for $B=0,0.5,1,1.5 \dots ,12$ T.

\begin{figure}[h]
\includegraphics{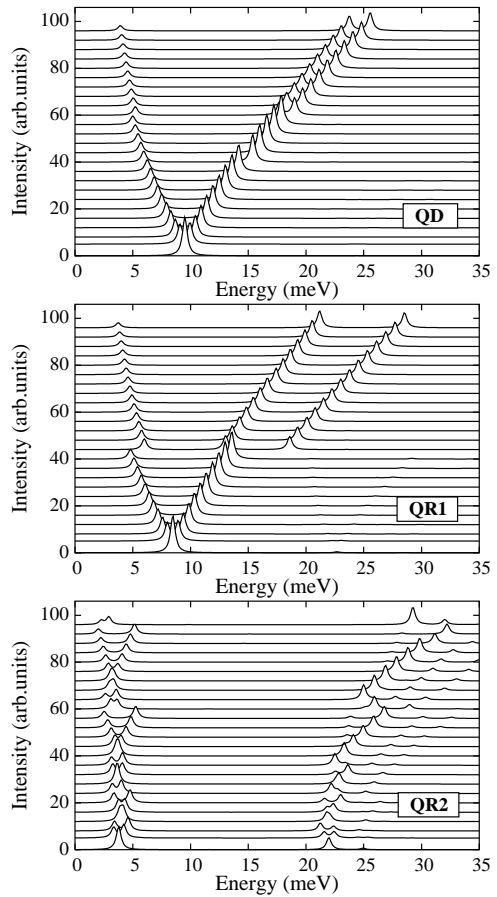}
\caption{FIR absorption of two electrons in QD (upper panel), QR1 (middle panel) and QR2 (lower panel) at $T=0$ K.
The spectra are calculated for magnetic fields $B=0-12$ T in steps of $0.5$ T. The intensities are in arbitrary units and the curves have been offset for clarity.}
\label{Figure6}
\end{figure}

It can be seen that the two-electron QD absorption is very similar to the single-electron one, except for a significantly higher intensity and a small ($<1.5$ meV) splitting of the $\Delta M=1$ branch above $7.5$ T.\cite{noKohn} This result was to be expected since self-assembled dots are well described by parabolic potentials, and hence the generalized Kohn theorem prevents many-electron effects from being revealed by excitation spectroscopy.\cite{Chakra_book}
Conversely, the hole of the rings breaks down the generalized Kohn theorem, so that the two-electron absorption spectra of QR1 and QR2 exhibit interesting differences with respect to the one-electron cases.
At low magnetic fields ($B \le 5.0$ T), the spectra of QR1 resemble those of the one-electron case. 
However, the ground state crossing at about $B=5.1$ T brings about a very different picture.
First, an abrupt shift in the $\Delta M=-1$ branch is seen at $B \ge 5.5$ T (which is a lower value of $B$ than in the one-electron case). Second, the $\Delta M=1$ branch splits into two parallel branches, similar to the QD case but with a
significantly larger energy spacing of about $5.5$ meV. One of the branches is a prolongation of the $\Delta M=1$ branch observed before $B=5.1$ T and the other one is abruptly shifted toward higher energies. The low-energy and high-energy $\Delta M=1$ branches originate from the $\Delta N=0$ and $\Delta N=1$ transitions, respectively. 
We would like to point out that this double $\Delta M=1$ branch feature was not found in previous calculations of low-barrier rings.\cite{PuentePRB}
The influence of the increased hole on the spectra of QR2 is even greater: both the low-lying $\Delta N=0$ set of resonances and the high-lying $\Delta N=1$ resonances are visible at $B=0$, and the rapid Aharonov-Bohm oscillations give rise to many small shifts in the energies of the resonances, as well as to an energy gap between 7 meV and 20 meV. The most visible effect of the electron-electron interaction on the FIR absorption of QR2 is an increase in the number of oscillation-induced energy shifts.

\begin{figure}[h]
\includegraphics{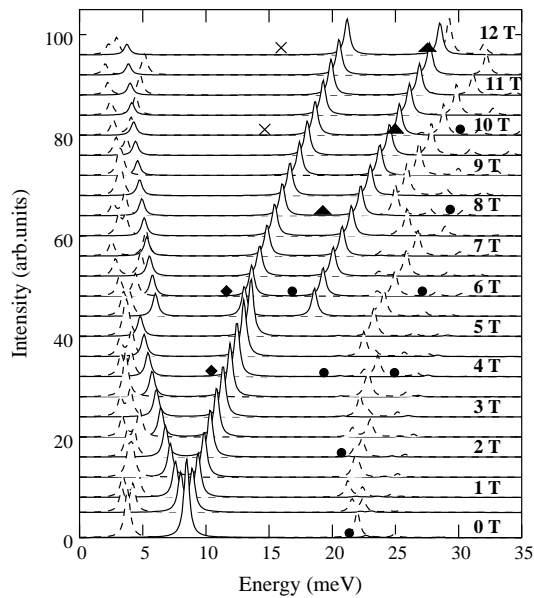}
\caption{Combined FIR absorption of two electrons in QR1 and QR2 at $T=0$ K.
The spectra are calculated for magnetic fields $B=0-12$ T in steps of $0.5$ T. The intensities are in arbitrary units and the curves have been offset for clarity. 
Solid lines are used for QR1 and dashed lines for QR2. The symbols represent the experimental resonances.}
\label{Figure7}
\end{figure}

Figure 7 illustrates the combined two-electron absorption spectra of QR1 and QR2. Dashed lines are used for QR2 and solid lines for QR1. The experimental data are displayed using the original symbols of Ref.\onlinecite{LorkePRL}. Due to the improved signal-to-noise ratio, a larger number of experimental resonances are available for the two-electron systems.
A remarkable agreement between our calculations and the experimental resonances is observed. The dots are qualitatively described by the absorption of QR2. An exception is the low-energy dot at $B=6$ T, which we do not assign to QR2 but to the high-lying $\Delta M=1$ branch of QR1. The triangles are also well described by the high-lying $\Delta M=1$ branch of QR1. Finally, the low-lying $\Delta M=1$ branch of QR1 accounts for the diamonds and the crosses. 
We point out that the intensities of the low-lying and high-lying sets of resonances in Figure 7 are of the same order of magnitude, as found experimentally.
Even if some experimental points are only qualitatively described by our calculations, it can be concluded that our model suggests an alternative assignment of experimental resonances to that of Ref.\onlinecite{LorkePRL}. 
This assignment is similar to that suggested by other authors.\cite{PuentePRB} However, we also offer an explanation for the experimental resonances marked with crosses, which were not understood to date.\cite{EmperadorPRB,LorkePRL,HuPRB,PuentePRB}.

We want to stress that if we assume a mixture of QD and QR2 we would not account for the triangles in Figure 7, since the high-lying $\Delta N=1, \Delta M=1$ branch of QD is too close in energy to the $\Delta N=0, \Delta M=1$ one (see Figure 6). Moreover, the two-electron results lead us to propose the crosses in the one-electron system (Figure 4) to be assigned to the $\Delta n=0, \Delta m=1$ branch of QR1, and the triangles to a $\Delta n=1, \Delta m=1$ branch which is not allowed at $T=0$ K, but would be allowed at finite temperature when the first excited level is partially populated.
As in Ref.\onlinecite{PuentePRB}, our calculations indicate that the characteristic spectral features found in the experiment at $B=8$ T are not due to an Aharanov-Bohm oscillation but to a crossing between resonances of two types of quantum rings. Therefore, the true signature of a quantum ring (with the geometry revealed by atomic force microscopy, QR2) in the FIR absorption experiments of Ref.\onlinecite{LorkePRL} is the presence of an intense resonance at about 20-23 meV and $B=0$. Such a resonance was theoretically predicted by previous works using three-dimensional models and describing similar structures to QR2.\cite{ClimentePRBee,VoskoboynikovPRB,PlanellesPRB}
We also find a very close agreement between the energy of this experimental resonance (the dot at $B=0$ T) and the high-energy resonance of QR2 in both the one-electron and the two-electron spectra. This fact points out that our model gives a correct estimate of the Coulomb energy, as opposed to two-dimensional models which overestimate it due to the missing vertical motion.\cite{ClimentePRBee} 

\section{Conclusions}

We have studied the FIR absorption of one and two electrons in a quantum dot, in a quantum ring with a small hole and in a quantum ring with the dimensions measured by atomic force microscopy.
Our calculations show that it is possible to reproduce the FIR absorption experiments on self-assembled rings in a magnetic field by using our realistic three-dimensional model and assuming a mixture of the two rings with different inner radii.
We provide an alternative assignment of the experimental resonances to that suggested in Refs.\onlinecite{EmperadorPRB,LorkePRL}. This is the first theoretical description which accounts for the entire set of experimental resonances.

\begin{acknowledgments}
Financial support from a MEC-FPU grant, MEC-DGI project CTQ2004-02315/BQU, and
UJI-Bancaixa project P1-B2002-01 are gratefully acknowledged.
\end{acknowledgments}


\begin{thebibliography}{35}
\bibitem{ButtikerPL} M. B\"uttiker, Y. Imry, and R. Landauer, Phys. Lett. {\bf 96A}, 365 (1983).
\bibitem{AronovRMP} A.G. Aronov, and Yu. V. Sharvin, Rev. Mod. Phys. {\bf 59}, 755 (1987).
\bibitem{ChakraPRB} T. Chakraborty, and P. Pietil\"ainen, Phys. Rev. B {\bf 50}, 8460 (1994).
\bibitem{WendlerPSSb} L. Wendler, and V.M. Fomin, Phys. Status Solidi (b) {\bf 191}, 409 (1995).
\bibitem{FuhrerNAT} A. Fuhrer, S. Luescher, T. Ihn, T. Heinzel, K. Ensslin, W. Wegscheider, and M. Bichler, Nature (London) {\bf 413}, 6858 (2001).
\bibitem{AharonovPR} Y. Aharonov, D. Bohm, Phys. Rev. {\bf 115}, 485 (1959).
\bibitem{OlariuRMP} S. Olariu, and II. Popescu, Rev. Mod. Phys. {\bf 57}, 339 (1985).
\bibitem{ChandrasekharPRL} V. Chandrasekhar, R.A. Webb, M.J. Brady, M.B. Ketchen, W.J. Gallagher, A. Kleinsasser, Phys. Rev. Lett. {\bf 67}, 3578 (1991).
\bibitem{Chakra_book} T. Chakraborty, \emph{Quantum dots}, (Elsevier Science B.V., Amsterdam, 1999).
\bibitem{GarciaAPLRazAPL} J.M. Garcia, G. Medeiros-Ribeiro, K. Schmidt, T. Ngo, J.L. Feng, A. Lorke, J. Kotthaus, and
P.M. Petroff, Appl. Phys. Lett. {\bf 71}, 2014 (1997); T. Raz, D. Ritter, and G. Bahir, Appl. Phys. Lett. {\bf 82}, 17
06 (2003).
\bibitem{BayerPRL} M. Bayer, M. Korkusi\'nski, P. Hawrylak, T. Gutbrod, M. Michel, and A. Forchel, Phys. Rev. Lett. {\bf 90}, 186801 (2003).
\bibitem{SkolnickPE} M.S. Skolnick, and D.J. Mowbray, Physica E (Amsterdam) {\bf 21}, 155 (2004).
\bibitem{WarburtonNAT} R.J. Warburton, C. Schaflein, D. Haft, F. Bickel, A. Lorke, K. Karrai, J.M. Garcia, W. Schoenfeld, and P.M. Petroff, Nature (London) {\bf 405}, 926 (2000).
\bibitem{PettersonPE} H. Petterson, R.J. Warburton, A. Lorke, K. Karrai, J.P. Kotthaus, J.M. Garcia, and P.M. Petroff, Physica E (Amsterdam) {\bf 6}, 510 (2000).
\bibitem{HaftPE} D. Haft, C. Schulhauser, A.O. Govorov, R.J. Warburton, K. Karrai, J.M. Garcia, W. Schoenfeld, and P.M. Petroff, Physica E (Amsterdam) {\bf 13}, 165 (2002).
\bibitem{EmperadorPRB} A. Emperador, M. Pi, M. Barranco, and A. Lorke, Phys. Rev. B {\bf 62}, 4573 (2000).
\bibitem{LorkePRL} A. Lorke, R.J. Luyken, A.O. Govorov, J.P. Kotthaus, J.M. Garcia, and P.M. Petroff, Phys. Rev. Lett. {\bf 84}, 2223 (2000).
\bibitem{HuPRB} H. Hu, J.L. Zhu, J.J. Xiong, Phys. Rev. B {\bf 62}, 16777 (2000).
\bibitem{BarkerPRB} J.A. Barker, R.J. Warburton, E.P. O'Reilly, Phys. Rev. B {\bf 69}, 035327 (2004).
\bibitem{PuentePRB} A. Puente, and Ll. Serra, Phys. Rev. B {\bf 63}, 125334 (2001).
\bibitem{WarburtonICPS} R. J. Warburton, B. Urbaszek, E. J. McGhee, C. Schulhauser, A. Hogele, K. Karrai, A.O. Govorov, J. A. Barker, B. D. Gerardot, P.M. Petroff, and J. M. Garcia, \emph{26$^{th}$ International Conferenence on the Physics of Semiconductors Proceedings}, (Institute of Physics, Edinburgh, 2002).
\bibitem{ClimentePRB} J.I. Climente, J. Planelles, and W. Jaskolski, Phys. Rev. B {\bf 68}, 075307 (2003).
\bibitem{ClimentePRBee} J.I. Climente, J. Planelles, and F. Rajadell, unpublished.
\bibitem{VoskoboynikovPRB} O. Voskoboynikov, Y. Li, H.M. Lu, C.F. Shih, and C.P. Lee, Phys. Rev. B {\bf 66}, 155306 (2002).
\bibitem{DownesJAP} J.R. Downes, D.A. Faux, and E.P. O'Reilly, J. Appl. Phys. {\bf 81}, 6700 (1997).
\bibitem{DaviesJAP} J.H. Davies, J. Appl. Phys. {\bf 84}, 1358 (1998).
\bibitem{LiJAP} S.S. Li, J.B. Xia, J. Appl. Phys. {\bf 89}, 3434 (2001).
\bibitem{VoskoboynikovPE} O. Voskoboynikov, C.P. Lee, Physica E {\bf 20}, 278 (2004).
\bibitem{Loehr_book} J.P. Loehr, \emph{Physics of strained quantum well lasers}, (Boston Kluwer Academic cop. 1998).
\bibitem{ClimenteXXmag} J.I. Climente, J. Planelles, and J.L. Movilla, Phys. Rev. B {\bf 70}, 081301(R) (2004).
\bibitem{spintransPRB} The $(0,0)$ symmetry window is even wider ($0-11$ T) in a ring with a slightly smaller
outer radius of 60 nm.\cite{ClimenteXXmag} This difference arises from the stronger confinement in the latter structure,
 which leads to larger energy spacing between electron states at $B=0$.
\bibitem{WagnerPRB} M. Wagner, U. Merkt, and A.V. Chaplik, Phys. Rev. B {\bf 45}, 1951 (1992).
\bibitem{noKohn} This splitting should not be observed in a perfect parabolic dot.
\bibitem{PlanellesPRB} J. Planelles, W. Jaskolski, and J.I. Aliaga, Phys. Rev. B {\bf 65}, 033306 (2002).

\end{thebibliography}
\end{document}